\documentclass[aps,pra,showpacs,preprint]{revtex4}
\usepackage{graphicx}
\usepackage{amsmath,amssymb,amsfonts}
\usepackage{dcolumn}
\usepackage[english]{babel}
\usepackage{epsfig}

\begin{document}
\title{The $J$-matrix inverse scattering approach for coupled channels with different
thresholds}
\author{ S. A. Zaytsev}
\email[E-mail: ]{zaytsev@fizika.khstu.ru} \affiliation{Pacific
National University, Khabarovsk, 680035, Russia}

\begin{abstract}
The inverse scattering method within the $J$-matrix approach to the
two coupled-channel problem is discussed. We propose a
generalization of the procedure to the case with different
thresholds.
\end{abstract}
 \maketitle

\section{Introduction}
In the $J$-matrix approach \cite{YF,BR} to the many-channel problem
the partial wave potentials $V^{(\alpha\,\beta)}$ in the interaction
\begin{equation}\label{Pot}
    \widehat{V}=\sum \limits _{\alpha, \, \beta}\left|
    \alpha\right> V^{(\alpha\,\beta)}\left<\beta\right|
\end{equation}
are given by the expansion
\begin{equation}\label{PartialPot}
    V^{(\alpha\,\beta)}(r,\,r')=\hbar\omega\sum \limits _{n, \,
    n'=0}^{N-1}\phi_n^{\ell_{\alpha}}(x)\,
    V_{n,n'}^{\alpha\,\beta}\,\phi_{n'}^{\ell_{\beta}}(x').
\end{equation}
Here, $\alpha$ is the channel index that contains the orbital
angular momentum $\ell_{\alpha}$, $x=r/\rho$, the relative
coordinate in units of the oscillator radius $\rho=\sqrt{\hbar/\mu
\omega}$, where $\mu$ is the reduced mass. In the case of collisions
with neutral targets, the harmonic oscillator basis functions
\begin{equation}
\phi_n^{\ell}(x)=(-1)^n\,\sqrt{\frac{2n!}{\rho
\Gamma(n+\ell+\frac32)}}\, x^{\ell+1}e^{-x^2/2}\,L_n^{\ell+1/2}(x^2)
\label{bf}
\end{equation}
are applied.

In \cite{TMP,JPG,PRCJ} an inverse scattering formalism within the
$J$-matrix method has been proposed and developed, where potentials
(\ref{Pot}), (\ref{PartialPot}) are determined from a given
$S$-matrix. In the previous version \cite{TMP,JPG,PRCJ} we have
restricted ourselves to the case of a two-channel system without a
threshold.

In this paper we attempt to extend the inverse scattering procedure
\cite{TMP,JPG,PRCJ} to the two coupled channels with different
thresholds: $\Delta_1=0$, $\Delta_2=\Delta>0$. The channel wave
numbers $k_{\alpha}$ are related by
\begin{equation}\label{k_alpha}
    k_{\alpha}^2=k^2-\Delta_{\alpha}.
\end{equation}
We measure the energy $E$ in $\hbar\omega$, i. e.
$E=\hbar\omega\epsilon$ and $\epsilon=q^2/2$, where $q=\rho k$.
Thus, the elements $H^{\alpha\,\beta}_{n\,m}$ of the Hamiltonian
$\mathcal{N}\times\mathcal{N}$ ($\mathcal{N}=2N$) matrix ${\bf H}$
in the basis
$\left\{\left|\alpha\right>\phi_n^{\ell_{\alpha}}\right.$,
$n=0,\,\ldots,\,N-1$, $\left.\alpha=1,\, 2 \right\}$ are written as
\begin{equation}\label{Hnm}
H^{\alpha\,\beta}_{n\,m}=\delta_{\alpha\,\beta}\left(T_{n\,m}^{(\alpha)}
                            +\delta_{n\,m}\frac{\rho^2}{2}\Delta_{\alpha}\right)
                            +V_{n,m}^{\alpha\,\beta}.
\end{equation}
Here, $T_{n\,m}^{(\alpha)}\equiv T_{n,\,m}^{\ell_{\alpha}}$ are the
elements of the symmetric tridigonal matrix
\begin{equation}
 \begin{array}{c}
T_{n,\,n}^{\ell}=\frac12\left(2n+\ell+\frac32\right),\\[3mm]
T_{n,\,n+1}^{\ell}=T_{n+1,\,n}^{\ell}=-\frac12\sqrt{(n+1)
\left(n+\ell+\frac32\right)}\\
 \end{array}
 \label{Tell}
\end{equation}
of the $\ell$th partial wave kinetic energy operator
\begin{equation}\label{Tr}
    \widehat{T}_{\ell}=\frac{\rho^2}{2}\left(-\frac{d^2}{dr^2}
    +\frac{\ell(\ell+1)}{r^2}\right).
\end{equation}

Recall that generally, with a finite order potential matrix it is
possible to reproduce the scattering data only in finite energy
interval. The $S$-matrix (${\bf S}(k)$) features in a given interval
$k \in [0, \, k_0]$ determine the optimal combination of $N$ and
$\rho$ that is best suited to the task of the potential (\ref{Pot}),
(\ref{PartialPot}) construction.

Within the framework of the method the eigenvalues $\left\{\lambda_j
\right\}$ and rows $\left\{Z_{N,j}\right\}$,
$\left\{Z_{\mathcal{N},j}\right\}$ of the eigenvector matrix ${\bf
Z}$ of the Hamiltonian matrix ${\bf H}$ are derived from the
$S$-matrix. As has been shown in Ref. \cite{PRCJ}, the set
$\left\{\lambda_j, \, Z_{N,j}, \, Z_{\mathcal{N},j}
\right\}_{j=1}^{\mathcal{N}}$ suffices to determine the Hamiltonian
matrix ${\bf H}$  of the quasitridiagonal form
\begin{equation}\label{Hmat}
{\bf H} = \left(
\begin{array}{ccccc|ccccc}
  a_0^{(1)} & b_1^{(1)} &  &  &  & u_0 &  &  &  &  \\
  b_1^{(1)} & a_1^{(1)} & b_2^{(1)} & & \mbox{\large 0}& v_1 & u_1 &  & \mbox{\large 0} &  \\
   & \ddots & \ddots & \ddots &  &  & \ddots & \ddots &  &  \\
   &  & b_{N-2}^{(1)} & a_{N-2}^{(1)} & b_{N-1}^{(1)} &  &  & v_{N-2} & u_{N-2} &  \\
   & \mbox{\large 0} &  & b_{N-1}^{(1)} & a_{N-1}^{(1)} & \mbox{\large 0} &  &  & v_{N-1} & u_{N-1}\\
 \hline
  u_0 & v_1 &  &  & \mbox{\large 0} & a_0^{(2)}
   & b_1^{(2)} &  & \mbox{\large 0} &  \\
   & u_1 & v_2 &  &  & b_1^{(2)} & a_1^{(2)}
 & b_2^{(2)} &  &  \\
   &  & \ddots & \ddots &  &  & \ddots & \ddots & \ddots &\\
   &  &  & u_{N-2} & v_{N-1} &  &  & b_{N-2}^{(2)} & a_{N-2}^{(2)}
& b_{N-1}^{(2)} \\
   & \mbox{\large 0} &  &  & u_{N-1}&  & \mbox{\large 0} &  & b_{N-1}^{(2)} &
   a_{N-1}^{(2)}\\
\end{array}
 \right).
\end{equation}

The present paper is organized as follows. In Sec.~II we outline the
multichannel $J$-matrix formalism. Sec.~III is devoted to the
generalization of the $J$-matrix inverse scattering formalism to the
case of two coupled channels in the presence of the threshold. An
example illustrated all steps of the presented inverse procedure is
given in Sec.~IV. Conclusions are drawn in Sec.~V. The Appendix
consists of a brief discussion of the $J$-matrix version of the
two-channel Marchenko equations.

\section{Preliminaries}
The components $\psi^{(\alpha\beta)}(k,\,r)$ of $2\times2$ matrix
solution to the coupled radial Schr\"{o}dinger equation, within the
$J$-matrix formalism, are expanded in the oscillator function series
\cite{BR}
\begin{equation}\label{psiab}
\psi^{(\alpha\beta)}(k,\,r)=\sum \limits _{n=0}^{\infty}
\phi_n^{\ell_{\alpha}}(x)c^{(\alpha\,\beta)}_n(k).
\end{equation}
For the potential (\ref{Pot}), (\ref{PartialPot}) the expansion
coefficients $c^{(\alpha\,\beta)}_n(k)$ are represented for $n\ge
N-1$ as a combination
\begin{equation}\label{pabas}
c^{(\alpha\,\beta)}_n(k)=f^{(\alpha\,\beta)}_n(k), \quad
f^{(\alpha\,\beta)}_n(k)\equiv\frac{i}{2}
\left[\;\mathcal{C}_{n,\,\ell_{\alpha}}^{(-)}(q_{\alpha})\delta_{\alpha\,\beta}
-\mathcal{C}_{n,\,\ell_{\alpha}}^{(+)}(q_{\alpha})\sqrt{\frac{q_{\beta}}{q_{\alpha}}}
\,S_{\alpha\,\beta}(k)\; \right]
\end{equation}
of two linearly independent solutions
\begin{equation}
 \begin{array}{c}
\mathcal{C}_{n,\,\ell}^{(\pm)}(q)=\mathcal{C}_{n,\,\ell}(q)
\pm i\mathcal{S}_{n,\,\ell}(q),\\[3mm]
\mathcal{S}_{n,\,\ell}(q) = \sqrt{\frac{\pi\rho
n!}{\Gamma(n+\ell+\frac32)}}\,
q^{\ell+1}e^{-q^2/2}\,L_n^{\ell+1/2}(q^2),\\[3mm]
\mathcal{C}_{n,\,\ell}(q) = \sqrt{\frac{\pi\rho
n!}{\Gamma(n+\ell+\frac32)}}\,
\frac{\Gamma(\ell+1/2)}{\pi\,q^{\ell}}\,e^{-q^2/2}\,
F\left(-n-\ell-1/2,\, -\ell+1/2;\, q^2 \right),\\
 \end{array}
 \label{SC}
\end{equation}
to the three-term recursion relation
\begin{equation}
T_{n, \, n-1}^{\ell}\,d_{n-1}(q)+ T_{n, \, n}^{\ell}\,d_n(q)+T_{n,
\, n+1}^{\ell}\,d_{n+1}(q)=\frac{q^2}{2} \,d_n(q), \quad n=1, \, 2,
\, \ldots \, ,\label{TRR}
\end{equation}
which is the basis-set representation of the free Schr\"{o}dinger
equation.

In turn, the component $\psi_{\nu}^{(\alpha\,\beta)}$ of the bound
eigenstate of energy $E_{\nu}=-\frac{\hbar^2\kappa_{\nu}^2}{2\mu}$
expansion
\begin{equation}\label{bs}
    \psi_{\nu}^{(\alpha\,\beta)}(r) = \sum \limits
    _{n=0}^{\infty}\phi_n^{\ell_{\alpha}}(x)c^{(\alpha\,\beta)}_n(\nu)
\end{equation}
coefficients satisfy the boundary conditions
\begin{equation}\label{bsas}
 \begin{array}{c}
c_n^{(\alpha\,\beta)}(\nu)=\delta_{\alpha\,\beta}f_n^{(\alpha)}(\nu),\quad
n\ge N-1,\\[3mm]
                  f_n^{(\alpha)}(\nu)\equiv i^{\ell_{\alpha}}
\,\mathcal{C}^{(+)}_{n,\ell_{\alpha}}
(i\rho\sqrt{\kappa_{\nu}^2+\Delta_{\alpha}}).\\
 \end{array}
\end{equation}

The completeness relation of the scattering and bound states
\cite{CS} can be transformed for the solutions
\begin{equation}\label{ccnu}
    {\bf c}_n(k) = \left(
                  \begin{array}{cc}
                    c_n^{(11)}(k) & c_n^{(12)}(k) \\
                    c_n^{(21)}(k) & c_n^{(22)}(k) \\
                  \end{array}
                \right),\qquad
    {\bf c}_n^{(\nu)} = \left(
                  \begin{array}{cc}
                    c_n^{(11)}(\nu) & c_n^{(12)}(\nu) \\
                    c_n^{(21)}(\nu) & c_n^{(22)}(\nu) \\
                  \end{array}
                \right)
\end{equation}
into
\begin{equation}\label{CR}
    \frac{2}{\pi}\int \limits_{0}^{\infty}dk\,{\bf
    c}_n(k)\,{\bf P}\,{{\bf c}_m(k)}^{\dag}+\sum \limits _{\nu}
  {\bf c}_n^{(\nu)}{\bf A}_{\nu}{{\bf c}_m^{(\nu)}}^{\dag}={\bf I}\delta_{n,m},
\end{equation}
${\bf I}$ being the $2\times2$ unit matrix. Here the matrices ${\bf
P}$ and ${\bf A}$ are defined as
\begin{equation}\label{PA}
    {\bf P}=\left(
              \begin{array}{cc}
                1 & 0 \\
                0 & \frac{k}{k_2}\frac{Re(k_2)}{\left|k_2\right|} \\
              \end{array}
            \right)\qquad
    {\bf A}_{\nu} = \left(
                  \begin{array}{cc}
                    \left(\mathcal{M}^{(1)}_{\nu}\right)^2 &
                    \mathcal{M}^{(1)}_{\nu}\mathcal{M}^{(2)}_{\nu} \\
                    \mathcal{M}^{(2)}_{\nu}\mathcal{M}^{(1)}_{\nu}
                    & \left(\mathcal{M}^{(2)}_{\nu}\right)^2 \\
                  \end{array}
                \right),
\end{equation}
where \cite{BZP}
\begin{equation}\label{SRes}
    i\mathop{Res}\limits _{k=
    i \kappa_{\nu}}S_{\alpha\,\beta}(k) = i^{\ell_{\alpha}+\ell_{\beta}}
    \sqrt{\frac{\sqrt{\kappa_{\nu}^2+\Delta_{\alpha}}\sqrt{\kappa_{\nu}^2+\Delta_{\beta}}}
    {\kappa_{\nu}^2}}\mathcal{M}^{(\alpha)}_{\nu}\mathcal{M}^{(\beta)}_{\nu}.
\end{equation}
$\mathcal{M}^{(\alpha)}_{\nu}$ are the asymptotic normalization
constants, i. e. the normalized bound-state wave function
\begin{equation}\label{BSWF}
    \left(
      \begin{array}{c}
        \varphi_{\nu}^{(1)} \\
        \varphi_{\nu}^{(2)} \\
      \end{array}
    \right)=\left(
              \begin{array}{cc}
                \psi_{\nu}^{(11)}(r) & \psi_{\nu}^{(12)}(r) \\
                \psi_{\nu}^{(21)}(r) & \psi_{\nu}^{(22)}(r) \\
              \end{array}
            \right)\left(
                     \begin{array}{c}
                       \mathcal{M}^{(1)}_{\nu}\\
                       \mathcal{M}^{(2)}_{\nu}\\
                     \end{array}
                   \right)
\end{equation}
asymptotically behaves as
\begin{equation}\label{ABS}
        \left(
      \begin{array}{c}
        \varphi_{\nu}^{(1)} \\
        \varphi_{\nu}^{(2)} \\
      \end{array}
    \right) \mathop{\sim}\limits _{r\rightarrow\infty}
    \left(
     \begin{array}{c}
      \mathcal{M}^{(1)}_{\nu}e^{-r\kappa_{\nu}} \\
      \mathcal{M}^{(2)}_{\nu}e^{-r\sqrt{\kappa_{\nu}^2+\Delta}} \\
     \end{array}
   \right).
\end{equation}

The $J$-matrix expressions for the $S$-matrix elements have the form
\begin{equation}\label{S11}
 \begin{array}{l}
    S_{11}(k) = \frac{1}{D^{(+)}(k)}\left[\left(\mathcal{C}_{N-1,\ell_1}^{(-)}(q_1)-
     \mathcal{P}_{11}(\epsilon)\,T^{(1)}_{N-1,N}\mathcal{C}_{N,\ell_1}^{(-)}(q_1)\right)
                    \right.
                    \times\\[3mm]
                    \times
                    \left.
                    \left(\mathcal{C}_{N-1,\ell_2}^{(+)}(q_2)-
     \mathcal{P}_{22}(\epsilon)\,T^{(2)}_{N-1,N}\mathcal{C}_{N,\ell_2}^{(+)}(q_2)\right)
                 -\mathcal{P}^2_{12}(\epsilon)\,T^{(1)}_{N-1,N}T^{(2)}_{N-1,N}
                 \mathcal{C}_{N,\ell_1}^{(-)}(q_1)\mathcal{C}_{N,\ell_2}^{(+)}(q_2)
                 \right],\\
 \end{array}
\end{equation}

\begin{equation}\label{S22}
 \begin{array}{l}
    S_{22}(k) = \frac{1}{D^{(+)}(k)}\left[\left(\mathcal{C}_{N-1,\ell_1}^{(+)}(q_1)-
     \mathcal{P}_{11}(\epsilon)\,T^{(1)}_{N-1,N}\mathcal{C}_{N,\ell_1}^{(+)}(q_1)\right)
                    \right.
                    \times\\[3mm]
                    \times
                    \left.
                    \left(\mathcal{C}_{N-1,\ell_2}^{(-)}(q_2)-
     \mathcal{P}_{22}(\epsilon)\,T^{(2)}_{N-1,N}\mathcal{C}_{N,\ell_2}^{(-)}(q_2)\right)
                 -\mathcal{P}^2_{12}(\epsilon)\,T^{(1)}_{N-1,N}T^{(2)}_{N-1,N}
                 \mathcal{C}_{N,\ell_1}^{(+)}(q_1)\mathcal{C}_{N,\ell_2}^{(-)}(q_2)
                 \right],\\
 \end{array}
\end{equation}

\begin{equation}\label{S12}
    S_{12}(k) = -\frac{i \rho^2\sqrt{k_1\,k_2}
    \mathcal{P}_{12}(\epsilon)}{D^{(+)}(k)},
\end{equation}
where
\begin{equation}\label{Dplus}
 \begin{array}{l}
    D^{(+)}(k) = \left(\mathcal{C}_{N-1,\ell_1}^{(+)}(q_1)-
     \mathcal{P}_{11}(\epsilon)\,T^{(1)}_{N-1,N}\mathcal{C}_{N,\ell_1}^{(+)}(q_1)\right)
                    \times\\[3mm]
                    \quad \times
                    \left(\mathcal{C}_{N-1,\ell_2}^{(+)}(q_2)-
     \mathcal{P}_{22}(\epsilon)\,T^{(2)}_{N-1,N}\mathcal{C}_{N,\ell_2}^{(+)}(q_2)\right)
                 -\mathcal{P}^2_{12}(\epsilon)\,T^{(1)}_{N-1,N}T^{(2)}_{N-1,N}
                 \mathcal{C}_{N,\ell_1}^{(+)}(q_1)\mathcal{C}_{N,\ell_2}^{(+)}(q_2).\\
 \end{array}
\end{equation}
The functions $\mathcal{P}_{\alpha\,\beta}(\epsilon)$ are defined by
\begin{equation}\label{Pab}
    \mathcal{P}_{11}(\epsilon)=\sum \limits
    _{j=1}^{\mathcal{N}}\frac{Z_{N,j}^2}{\epsilon-\lambda_j},\quad
    \mathcal{P}_{12}(\epsilon)=\sum \limits
    _{j=1}^{\mathcal{N}}\frac{Z_{N,j}Z_{\mathcal{N},j}}{\epsilon-\lambda_j},
    \quad
    \mathcal{P}_{22}(\epsilon)=\sum \limits
    _{j=1}^{\mathcal{N}}\frac{Z_{\mathcal{N},j}^2}{\epsilon-\lambda_j}.
\end{equation}
The eigenvalues $\lambda_j$ are thus the poles of
$\mathcal{P}_{\alpha\,\beta}(\epsilon)$.

\section{Description of the method}
\subsection*{Region $\sqrt{\Delta}<k<k_0$}
In this energy region all channels are open, the $S$-matrix is
unitary. Thus the method \cite{TMP,PRCJ} can be applied to calculate
the eigenvalues $\lambda_j \in [\frac{\rho^2}{2}\Delta, \,
\frac{\rho^2}{2}k_0^2]$ and corresponding eigenvector components
$Z_{N,j}$, $Z_{\mathcal{N},j}$. Firstly, the functions
$\widetilde{\mathcal{P}}_{\alpha\,\beta}(\epsilon)$ are defined by
inverting (\ref{S11})-(\ref{Dplus}) relative to
$\mathcal{P}_{\alpha\,\beta}(\epsilon)$,
\begin{equation}\label{Ptab}
    \widetilde{\mathcal{P}}_{11}(\epsilon)=\frac{\Theta_1(\epsilon)}{D(\epsilon)},\quad
    \widetilde{\mathcal{P}}_{22}(\epsilon)=\frac{\Theta_2(\epsilon)}{D(\epsilon)},\quad
    \widetilde{\mathcal{P}}_{12}(\epsilon)=\frac{\Theta_3(\epsilon)}{D(\epsilon)},
\end{equation}
where
\begin{equation}\label{Theta1}
 \begin{array}{c}
    \Theta_1(\epsilon) = \left[
                    \left(\mathcal{C}_{N-1,\ell_1}^{(-)}(q_1)-
                          \mathcal{C}_{N-1,\ell_1}^{(+)}(q_1)\,S_{11}(k)\right)
                    \left(\mathcal{C}_{N,\ell_2}^{(-)}(q_2)-
                          \mathcal{C}_{N,\ell_2}^{(+)}(q_2)\,S_{22}(k)\right)-
                          \qquad  \right.\\[3mm]
                          \qquad \qquad \qquad \qquad \qquad
                          \qquad \qquad \qquad
                          -\left.\mathcal{C}_{N-1,\ell_1}^{(+)}(q_1)
                           \mathcal{C}_{N,\ell_2}^{(+)}(q_2)\,S_{12}^2(k)
                            \right]/T^{(1)}_{N-1,N},\\
 \end{array}
\end{equation}

\begin{equation}\label{Theta2}
 \begin{array}{c}
    \Theta_2(\epsilon) = \left[
                    \left(\mathcal{C}_{N,\ell_1}^{(-)}(q_1)-
                          \mathcal{C}_{N,\ell_1}^{(+)}(q_1)\,S_{11}(k)\right)
                    \left(\mathcal{C}_{N-1,\ell_2}^{(-)}(q_2)-
                          \mathcal{C}_{N-1,\ell_2}^{(+)}(q_2)\,S_{22}(k)\right)-
                          \qquad  \right.\\[3mm]
                          \qquad \qquad \qquad \qquad \qquad
                          \qquad \qquad \qquad
                          -\left.\mathcal{C}_{N,\ell_1}^{(+)}(q_1)
                                 \mathcal{C}_{N-1,\ell_2}^{(+)}(q_2)\,S_{12}^2(k)
                            \right]/T^{(2)}_{N-1,N},\\
 \end{array}
\end{equation}

\begin{equation}\label{Theta3}
 \Theta_3(\epsilon) = \frac{\mbox{i}\rho^2\,\sqrt{k_1k_2}\,S_{12}(k)}
                      {T^{(1)}_{N-1,N}\,T^{(2)}_{N-1,N}},
\end{equation}

\begin{equation}\label{D}
 \begin{array}{c}
    D(\epsilon) = \left(\mathcal{C}_{N,\ell_1}^{(-)}(q_1)-
                          \mathcal{C}_{N,\ell_1}^{(+)}(q_1)\,S_{11}(k)\right)
                    \left(\mathcal{C}_{N,\ell_2}^{(-)}(q_2)-
                          \mathcal{C}_{N,\ell_2}^{(+)}(q_2)\,S_{22}(k)\right)-
                          \qquad \qquad \qquad\\[3mm]
                          \qquad \qquad \qquad \qquad \qquad
                          \qquad \qquad \qquad \qquad \qquad
                          -\mathcal{C}_{N,\ell_1}^{(+)}(q_1)
                           \mathcal{C}_{N,\ell_2}^{(+)}(q_2)\,S_{12}^2(k).\\
 \end{array}
\end{equation}
Then the eigenvalues $\lambda_j$ are the roots of
\begin{equation}\label{Poles}
    D(\epsilon)=0.
\end{equation}
In turn the residues of
$\widetilde{\mathcal{P}}_{\alpha\,\beta}(\epsilon)$ (\ref{Ptab}) at
the poles $\lambda_j$ determine $Z_{N,j}$, $Z_{\mathcal{N},j}$
squared and the component products:
\begin{equation}\label{ZNj2}
    Z_{N,j}^2 =\mathop{Res}\limits _{\epsilon=\lambda_j}
    \widetilde{\mathcal{P}}_{11}(\epsilon),
\end{equation}

\begin{equation}\label{ZCNj2}
    Z_{\mathcal{N},j}^2 =\mathop{Res}\limits _{\epsilon=\lambda_j}
    \widetilde{\mathcal{P}}_{22}(\epsilon),
\end{equation}

\begin{equation}\label{ZCNNj}
    Z_{N,j}\,Z_{\mathcal{N},j} =\mathop{Res}\limits _{\epsilon=\lambda_j}
    \widetilde{\mathcal{P}}_{12}(\epsilon).
\end{equation}
Notice that the unitarity of the $S$-matrix provides that (for $N$
sufficiently large) the poles and residues of the functions
$\widetilde{\mathcal{P}}_{\alpha\,\beta}(\epsilon)$ (\ref{Ptab}) are
real:
\begin{equation}\label{ReR}
 Im(\lambda_j)=0,\qquad Im\left(\mathop{Res}\limits _{\epsilon=\lambda_j}
    \widetilde{\mathcal{P}}_{\alpha\,\beta}(\epsilon)\right)=0,
\end{equation}
and
\begin{equation}\label{Resaa}
    \mathop{Res}\limits _{\epsilon=\lambda_j}
    \widetilde{\mathcal{P}}_{\alpha\,\alpha}(\epsilon)\ge 0.
\end{equation}

\subsection*{Region $k<\sqrt{\Delta}$}
To define the functions $\widetilde{\mathcal{P}}_{\alpha\,\beta}$
(\ref{Ptab}) for this energies an analytic continuation of the
$S$-matrix to $k \in [0, \, \sqrt{\Delta}]$ would be carried out. To
our knowledge there is no general method for doing this. However,
the $S$-matrix couldn't behave here in an arbitrary way. Generally,
the use of non-unitary $S$-matrix can violate the conditions
(\ref{ReR}), (\ref{Resaa}). One way to meet the conditions is to
restrict oneself to the using of the only open-channel (unitary)
submatrix of the $S$-matrix. By this is meant that
Eqs.~(\ref{Ptab})-(\ref{D}) reduce to
\begin{equation}\label{Pt1}
    \widetilde{\mathcal{P}}_{1}(\epsilon)
    =\frac{\widetilde{\Theta}_1(\epsilon)}{\widetilde{D}(\epsilon)},
\end{equation}
\begin{equation}\label{Theta1t}
   \widetilde{\Theta}_1(\epsilon) =
                    \left(\mathcal{C}_{N-1,\ell_1}^{(-)}(q_1)-
                          \mathcal{C}_{N-1,\ell_1}^{(+)}(q_1)\,S_{11}(k)\right)
                            /T^{(1)}_{N-1,N},
\end{equation}

\begin{equation}\label{Dt}
    \widetilde{D}(\epsilon) = \mathcal{C}_{N,\ell_1}^{(-)}(q_1)-
                          \mathcal{C}_{N,\ell_1}^{(+)}(q_1)\,S_{11}(k).
\end{equation}
Then the eigenvalues $\lambda_j< \frac{\rho^2}{2}\Delta$ and the
corresponding components $Z_{N,j}$ of the eigenvectors are found
from
\begin{equation}\label{Pole1}
\widetilde{D}(\epsilon)=0
\end{equation}
and
\begin{equation}\label{Res1}
Z_{N,j}^2=\mathop{Res}\limits
_{\epsilon=\lambda_j}\widetilde{\mathcal{P}}_{1}(\epsilon).
\end{equation}
Finally, the components $Z_{\mathcal{N},j}$ are  assumed to be zero.

\subsection*{Region $k>k_0$}
As noted before \cite{PRCJ}, generally, a potential (\ref{Pot}),
(\ref{PartialPot}) of finite rank can describe the scattering data
only in a finite energy interval. In other words, the open-channel
submatrix of the resulting $S$-matrix coincides with the reference
one on the interval $[0, \, k_0]$, in the region $k>k_0$, however,
such is not the case. On the other hand, it is necessary that there
exist eigenvalues $\lambda_j>\frac{\rho^2}{2}k_0^2$. Since we have
no prior knowledge of the $S$-matrix for $k>k_0$, we couldn't draw
on the Eqs.~(\ref{Poles})-(\ref{ZCNNj}) to evaluate the ``external''
parameters $\left\{\lambda_j, Z_{N,j}, Z_{\mathcal{N},j} \right\}$,
$\lambda_j>\frac{\rho^2}{2}k_0^2$.

In the previous papers (see, e. g. \cite{PRCJ}) the ``external''
parameters are obtained through the use of a standard fit to the
data on the interval $k \in [0, \, k_0]$. In doing this the
constraints
\begin{equation}\label{OrthNorm}
    \sum \limits _{j=1}^{\mathcal{N}}Z_{N,j}^2=1,\quad
    \sum \limits _{j=1}^{\mathcal{N}}Z_{\mathcal{N},j}^2=1,\quad
    \sum \limits _{j=1}^{\mathcal{N}}Z_{N,j}Z_{\mathcal{N},j}=0,
\end{equation}
that follow from orthogonality of the eigenvector matrix ${\bf Z}$,
are allowed for the parameters $Z_{N,j}$, $Z_{\mathcal{N},j}$. In
the paper to this end we use the discrete version of the Marchenko
equations \cite{IPJ} generalized to the two coupled-channel case
(see Appendix). Let us assume that, as in an example discussed
below, only two largest eigenvalues $\lambda_j$ lie to the right of
$\frac{\rho^2}{2}k_0^2$. A set of six equations, that the
``external'' parameters $\left\{\lambda_j, Z_{N,j},
Z_{\mathcal{N},j} \right\}$, $j=N-1, \, N$ are found from, must
contain (\ref{OrthNorm}). In addition to this, equations can be
derived using relations between the elements of the Hamiltonian
matrix (\ref{Hmat}) and its eigenvalues and eigenvectors. In
particular,
\begin{equation}\label{a12uN}
    a^{(1)}_{N-1}=\sum
    \limits_{j=1}^{\mathcal{N}}\lambda_j\,Z_{N,j}^2,\quad
    a^{(2)}_{N-1}=\sum
    \limits_{j=1}^{\mathcal{N}}\lambda_j\,Z_{\mathcal{N},j}^2,\quad
    u_{N-1}=\sum
    \limits_{j=1}^{\mathcal{N}}\lambda_j\,Z_{N,j}\,Z_{\mathcal{N},j}.
\end{equation}
To obtain the quantities $a^{(1)}_{N-1}$, $a^{(2)}_{N-1}$ and
$u_{N-1}$ the Marchenko equations (\ref{MEQ}), (\ref{Knn}) are
solved for $K_{n\,m}^{(\alpha\,\beta)}$, $n=N-2$. Then
Eqs.~(\ref{a1n})-(\ref{vn}) are used.

Before we can use the Marchenko equations, we need to define the
matrix ${\bf S}(k)$ in the integrands on the right-hand side of
(\ref{Qab}) for $0 \le k < \infty$. For this purpose suppose that
there exists $k_{max}>0$ such that
\begin{equation}\label{Skm}
    {\bf S}(k)={\bf I}, \quad k>k_{max}.
\end{equation}
Notice that the integrands in Eq.~(\ref{Qab}) contain the solutions
$\mathcal{C}_{n,\,\ell}(q)$ which are exponentially large as $q
\rightarrow \infty$ \cite{IPJ}:
\begin{equation}\label{AC}
  \mathcal{C}_{n\,\ell}(q)\mathop{\sim}\limits_{q\rightarrow\infty}
 (-1)^{n+1}\sqrt{\rho\,n!\;\Gamma(n+\ell+3/2)/\pi}\;
  q^{-(2n+\ell+2)}e^{q^2/2}.
\end{equation}
Thus the assumption (\ref{Skm}) provides the convergence of the
integrals in (\ref{Qab}). On the other hand, since the goal is to
reproduce the scattering data in the interval $k \in [0, \, k_0]$,
it is natural to assume $k_{max}=k_0$. Clearly, the open-channel
submatrix of the matrix ${\bf S}(k)$ in the interval $[0, \, k_0]$
is determined by the scattering data.

We also define the matrix ${\bf S}(k)$ in Eq.~(\ref{Qab}) for
energies where the second channel is closed. Notice that because of
projection onto open channels (the matrix ${\bf P}$ in
Eq.~(\ref{Qab})) the contribution from the closed channel to the
kernels $Q_{n\,m}^{(\alpha\,\beta)}$ reduces to the integrals over
the interval $[0, \, \sqrt{\Delta}]$ of the terms containing
$f_n^{(21)}(k) =
-\frac{i}{2}\mathcal{C}_{n,\,\ell_2}^{(+)}(q_2)\sqrt{\frac{k}{k_2}}
\,S_{1\,2}(k)$. In turn, for large $n$ the solution
$\mathcal{C}_{n,\,\ell}^{(+)}(i|q|)$ behaves as \cite{TMF2}
\begin{equation}\label{CPlusi}
\mathcal{C}_{n,\,\ell}^{(+)}(i|q|) \sim
i^{-\ell}\,\sqrt{\rho}\,\left(n+\ell/2+3/4\right)^{-\frac14}\,
    e^{-|q|\sqrt{n+\frac{\ell}{2}+\frac{3}{4}}}.
\end{equation}
Hence choosing $N$ large enough, the integrated terms involving
$f_n^{(21)}(k)$ ($n \ge N-2$) for $k < \sqrt{\Delta}$ can be made as
small as we please. Thus for $N$ sufficiently large the
approximation
\begin{equation}\label{S12in}
S_{12}(k)=0, \quad k<\sqrt{\Delta}
\end{equation}
is acceptable for the matrix ${\bf S}(k)$ in Eq.~(\ref{Qab}).

Let ${\bf S}^{(0)}(k)$ denote the $S$-matrix corresponding to the
Hamiltonian matrix (\ref{Hmat}) obtained by using the approximations
(\ref{Skm}), (\ref{S12in}). Clearly, the properties (\ref{Skm}),
(\ref{S12in}) of the matrix ${\bf S}(k)$ are not shared by the
resulting matrix ${\bf S}^{(0)}(k)$. Thus, the following iteration
procedure allow us to evaluate a contribution from the closed
channel. At each iteration ($i$), in the integrands of the integrals
over the interval $[0, \, \sqrt{\Delta}]$ on the right side of
(\ref{Qab}) that contain $S_{1\,2}(k)$ the element of the matrix
${\bf S}^{(i-1)}(k)$ obtained at the previous step is used. The
procedure is carried out until convergence is achieved.

\subsection*{Bound states}
To every bound state with $\kappa_{\nu}$, $\mathcal{M}_{\nu}^{(1)}$,
$\mathcal{M}_{\nu}^{(2)}$ there corresponds the triplet $\left\{
\lambda_{\nu}, \, Z_{N,\, \nu}, \, Z_{\mathcal{N},\,\nu}\right\}$.
The ``bound'' parameters $\left\{\lambda_{\nu},\, Z_{N,\, \nu},\,
Z_{\mathcal{N},\,\nu}\right\}$ and $\left\{\kappa_{\nu},\,
\mathcal{M}_{\nu}^{(1)},\,\mathcal{M}_{\nu}^{(2)}\right\}$ are
related by
\begin{equation}\label{Dik}
    D^{(+)}(i\kappa_{\nu})=0
\end{equation}
and Eqs.~(\ref{SRes}). Notice that $\lambda_{\nu}$, $Z_{N,\, \nu}$,
$Z_{\mathcal{N},\,\nu}$ appear in the left members of Eqs.
(\ref{Dik}) and (\ref{SRes}) through the functions
$\mathcal{P}_{\alpha\,\beta}$ (\ref{Pab}). In turn, $\kappa_{\nu}$,
$\mathcal{M}_{\nu}^{(1)}$, $\mathcal{M}_{\nu}^{(2)}$ are involved in
the kernels $Q_{n\,m}^{(\alpha\,\beta)}$ (\ref{Qab}) of the
Marchenko equations. Thus all of the unknown parameters $\left\{
\lambda_j, \, Z_{N,\, j}, \, Z_{\mathcal{N},\,j}\right\}$,
$\lambda_j>\frac{\rho^2}{2}\Delta$ and $\left\{ \lambda_{\nu}, \,
Z_{N,\, \nu}, \, Z_{\mathcal{N},\,\nu}\right\}$ are found by solving
the system that consists of (\ref{OrthNorm}), (\ref{a12uN}),
(\ref{Dik}) and two out of four Eqs. (\ref{SRes}).

\section{Example}
As an example, we consider the $S$-matrix
\begin{equation}\label{Sab}
 \begin{array}{c}
      S_{11} = \frac{\displaystyle (x-i\,k)(a^2-b^2+i\,a\,k-i\,a\,k_2+k\,k_2)}
      {\displaystyle (x+i\,k)(a^2-b^2-i\,a\,k-i\,a\,k_2-k\,k_2)},\\[3mm]
      S_{12} = \frac{\displaystyle -2ib\,\sqrt{kk_2}(\sqrt{x^2+\Delta^2}-i\,k_2)}
      {\displaystyle (x+i\,k)(a^2-b^2-i\,a\,k-i\,a\,k_2-k\,k_2)},\\[3mm]
      S_{22} = \frac{\displaystyle (\sqrt{x^2+\Delta^2}-i\,k_2)(a^2-b^2-i\,a\,k+i\,a\,k_2+k\,k_2)}
                    {\displaystyle (\sqrt{x^2+\Delta^2}+i\,k_2)(a^2-b^2-i\,a\,k-i\,a\,k_2-k\,k_2)},\\
 \end{array}
\end{equation}
corresponding to a model $s$-wave $2\times2$ potential discussed in
Ref. \cite{SSB} with $a = -2$, $b = 0.6$, $x = 3$. A threshold
energy $\Delta=10$ is assumed in the second channel. (We take
$\hbar=\mu=1$.) We found that the number $N=5$ of the basis
functions (\ref{bf}) used in the expansion (\ref{PartialPot}) and
the oscillator radius quantity $\rho=0.495$ are best suited to the
task of a potential (\ref{Pot}), (\ref{PartialPot}) construction
that reproduces the $S$-matrix (\ref{Sab}) in the interval $k \in
[0,\, k_0]$, $k_0 = 6$.

The eigenvalues $\lambda_j$, $j=4,\ldots,8$ lie in the energy
interval $[\frac{\rho^2}{2}\Delta, \frac{\rho^2k_0^2}{2}]$ where the
second channel is open. Then this eigenvalues and the corresponding
eigenvector components are calculated from
(\ref{ZNj2})-(\ref{ZCNNj}). The results are displayed in Table~I.

The eigenvalues $\lambda_j \in [0, \, \frac{\rho^2}{2}\Delta]$,
$j=2,\,3$ (see Table~I) and the corresponding components $Z_{N,j}$,
$Z_{\mathcal{N},j}$ are found from the approximate
Eqs.~(\ref{Pole1})-(\ref{Res1}). Notice that if we use the
expression (\ref{Sab}) for the $S$-matrix in (\ref{D}), we obtain
that the equation (\ref{Poles}) has not real roots
$\lambda_j<\frac{\rho^2}{2}\Delta$.

The calculations of the eigenvalues
$\lambda_j>\frac{\rho^2k_0^2}{2}$ and the corresponding eigenvector
components are carried out in combination with the description of
the bound state that the system sustains at $E_b=-\kappa^2/2$,
$\kappa = 2.1946752413$. The corresponding residues of the
$S$-matrix elements are
\begin{equation}\label{SRess}
 \begin{array}{c}
    \mathop{Res}\limits_{k=i\kappa}S_{11}(k)=-i
    26.7100700336,\\[3mm]
    \mathop{Res}\limits_{k=i\kappa}S_{12}(k)= i 18.1352046367.\\
 \end{array}
\end{equation}
The ``bound'' parameters $\left\{\lambda_1, \, Z_{N,1}, \,
Z_{\mathcal{N},1} \right\}$ and the ``external'' ones
$\left\{\lambda_j, \, Z_{N,j}, \, Z_{\mathcal{N},j} \right\}$, $j=
9,\, 10$ are found by solving the system that consists of
(\ref{OrthNorm}), (\ref{a12uN}), (\ref{Dik}), (\ref{SRess}). In the
calculations of $a^{(1)}_{N-1}$, $a^{(2)}_{N-1}$, $u_{N-1}$ in
Eq.~(\ref{a12uN}) the approximations (\ref{Skm}), (\ref{S12in}) have
been applied. The resulting parameters $\left\{\lambda_j, \,
Z_{N,j}, \, Z_{\mathcal{N},j} \right\}$ are presented in Table~I
(a). Table~III(a) lists the corresponding Hamiltonian matrix
(\ref{Hmat}). In Figs.~1-3 are shown the eigenphase shifts
$\delta_1$, $\delta_2$ and the mixing parameter $\varepsilon$
corresponding to the matrix ${\bf S}(k)$ in the integrands on the
right-hand side of (\ref{Qab}) (thin solid line) and the resulting
matrix ${\bf S}^{(0)}(k)$ (dashed line).

In order to evaluate the effect of the closed channel the iteration
procedure described in the previous section is applied. The
convergence of $a^{(1)}_{N-1}$, $a^{(2)}_{N-1}$, $u_{N-1}$ on the
left-hand side of (\ref{a12uN}) depending on the number of
iterations is displayed in Table~II. The resulting parameters
$\left\{\lambda_j, \, Z_{N,j}, \, Z_{\mathcal{N},j} \right\}$,
($j=1,\,9,\,10$) and the Hamiltonian matrix (\ref{Hmat}) after 5
steps are listed in Tables~I(b) and Table~III(b), respectively. The
eigenphase shifts $\delta_1$, $\delta_2$ and the mixing parameter
$\varepsilon$ corresponding to the matrix ${\bf S}^{(5)}(k)$ are
plotted in Figs.~1-3 (dotted line).

A comparison of (a) and (b) results shows that, as we might expect,
the off-diagonal part $V^{(12)}=V^{(21)}$ of the interaction
(\ref{Pot}), (\ref{PartialPot}) is most sensitive to the used
approximations. The mixing parameter $\varepsilon$ behavior (see
Figs.~3) reflects this dependence. Notice that the difference in
the mixing parameters (in the interval $[0, \, k_0]$) is within the
accuracy of the method.

\section{Conclusion}
A generalization of the $J$-matrix inverse scattering approach to
the case of two coupled channels with different threshold energies
has been discussed. All the modification introduced in the inverse
scheme \cite{TMP,PRCJ} (employed in the case of a two-channel system
without threshold) appear to be plane and relatively obvious. An
inverse procedure is proposed which focuss on reproducing the
scattering data in a given energy interval. On the other hand, the
procedure allows us to evaluate the contribution from the closed
channel to the sought-for potential.


\appendix{}
\section{The $J$-matrix version of the Marchenko equations}
$f_n^{(\alpha\,\beta)}$ (as well as
$\mathcal{C}_{n\,\ell_{\alpha}}^{(\pm)}$) satisfies the three-term
recursion relations (\ref{TRR}). It follows herefrom, in view of the
boundary conditions (\ref{pabas}), (\ref{bsas}) and the Hamiltonian
matrix quasitridiagonal form (\ref{Hmat}), that the coefficients
$c_n^{(\alpha\beta)}(k)$, $c_n^{(\alpha\beta)}(\nu)$ can be
expressed as
\begin{equation}\label{TRIEXP}
    c_n^{(\alpha\,\beta)}=\sum \limits _{\substack{m=n \\
    \gamma=1,2}}^{2N-n-2}K_{n\,m}^{(\alpha\,\gamma)}f_m^{(\gamma\,\beta)}.
\end{equation}
Besides, (\ref{pabas}), (\ref{bsas}) imply that
$K_{N-1,N-1}^{(\alpha\,\beta)}=\delta_{\alpha\,\beta}$. Further,
from the expansion (\ref{TRIEXP}) and the completeness relations
(\ref{CR}) it follows that ${\bf c}_n$ is orthogonal to every ${\bf
f}_m$ for $n<m$, i. e.
\begin{equation}\label{ORTH}
    \frac{2}{\pi}\int \limits_{0}^{\infty}dk\,{\bf
    c}_n(k)\,{\bf P}\,{{\bf f}_m(k)}^{\dag}+\sum \limits _{\nu}
  {\bf c}_n^{(\nu)}{\bf A}_{\nu}{{\bf f}_m^{(\nu)}}^{\dag}={\bf 0},
  \qquad m>n,
\end{equation}
where
\begin{equation}\label{fM}
    {\bf f}_n(k) = \left(
                  \begin{array}{cc}
                    f_n^{(11)}(k) & f_n^{(12)}(k) \\
                    f_n^{(21)}(k) & f_n^{(22)}(k) \\
                  \end{array}
                \right),\qquad
    {\bf f}_n^{(\nu)} = \left(
                  \begin{array}{cc}
                f_n^{(1)}(\nu)& 0 \\
                0 & f_n^{(2)}(\nu)\\
                  \end{array}
                \right).
\end{equation}
Inserting the expansion (\ref{TRIEXP}) into (\ref{ORTH}) and
(\ref{CR}) then yields
\begin{equation}\label{MarchenkoEQ}
 \sum \limits
_{n'=n}^{2N-n-2}{\bf K}_{n\,n'}{\bf Q}_{n'\,m}={\bf 0},\quad m>n,
\end{equation}
\begin{equation}\label{Complete1}
\sum \limits _{m=n}^{2N-n-2}{\bf K}_{n\,m}{\bf
Q}_{m\,n}\widetilde{\bf K}_{n\,n}={\bf I},
\end{equation}
where
\begin{equation}\label{Qab}
    {\bf Q}_{n\,m}=\frac{2}{\pi}\int \limits_{0}^{\infty}dk\,{\bf
    f}_n(k)\,{\bf P}\,{{\bf f}_m(k)}^{\dag}+\sum \limits _{\nu}
  {\bf f}_n^{(\nu)}{\bf A}_{\nu}{{\bf f}_m^{(\nu)}}^{\dag},
\end{equation}
\begin{equation}\label{KM}
 {\bf K}_{n\,m}=\left(
                  \begin{array}{cc}
                    K^{(11)}_{n\,m} & K^{(12)}_{n\,m} \\
                    K^{(21)}_{n\,m} & K^{(22)}_{n\,m} \\
                  \end{array}
                \right).
\end{equation}
Setting
\begin{equation}\label{MM}
    {\bf K}_{n\,m}={\bf K}_{n\,n}{\bf M}_{n\,m}, \quad m>n,
\end{equation}
we can rewrite (\ref{MarchenkoEQ}) in the form
\begin{equation}\label{MEQ}
\sum \limits _{n'=n+1}^{2N-n-2}{\bf M}_{n\,n'}{\bf Q}_{n'\,m} =
   -{\bf Q}_{n\,m},\quad m>n.
\end{equation}
Inserting (\ref{MM}) in (\ref{Complete1}) gives
\begin{equation}\label{Knn}
\widetilde{\bf K}_{n\,n}{\bf K}_{n\,n} = \left[{\bf Q}_{n\,n}+
  \sum \limits _{m=n+1}^{2N-n-2}{\bf M}_{n\,m}{\bf Q}_{m\,n} \right]^{-1}.
\end{equation}
Once ${\bf M}_{n\,m}$ have been found by solving (\ref{MEQ}), ${\bf
K}_{n\,n}$ can be evaluated from (\ref{Knn}). Notice that in this
case the Hamiltonian matrix quasitridiagonal form (\ref{Hmat})
implies that $K_{n\,n}^{(21)}=0$.

A knowledge of $K_{n\,m}^{(\alpha\, \beta)}$ finally furnishes the
Hamiltonian matrix (\ref{Hmat}) according to
\begin{equation}\label{a1n}
 \begin{array}{c}
  a_n^{(1)} =
  T_{n\,n}^{(1)}+\frac{K^{(11)}_{n\,n+1}}{K^{(11)}_{n\,n}}T_{n+1,\,n}^{(1)}
  -\frac{K^{(12)}_{n,\,n}\,K^{(21)}_{n-1,\,n}}
  {K^{(11)}_{n,\,n}K^{(22)}_{n-1,\,n-1}}T_{n,\,n-1}^{(2)}-\qquad \qquad \qquad \qquad \\[4mm]
  \qquad \qquad \qquad \qquad \qquad
  -\left( \frac{K^{(11)}_{n-1,\,n}}{K^{(11)}_{n-1,\,n-1}}
  -\frac{K^{(12)}_{n-1,\,n-1}K^{(21)}_{n-1,\,n}}
  {K^{(11)}_{n-1,\,n-1}\,K^{(22)}_{n-1,\,n-1}}\right)T_{n,\,n-1}^{(1)},
  \end{array}
\end{equation}
\begin{equation}\label{b1n}
  b^{(1)}_n = \frac{K^{(11)}_{n,\,n}}{K^{(11)}_{n-1,\,n-1}}T_{n,\,n-1}^{(1)},
\end{equation}
\begin{equation}\label{a2n}
 \begin{array}{c}
  a_n^{(2)} =
  T_{n\,n}^{(2)}+\frac{\rho^2}{2}\Delta+\frac{K^{(22)}_{n\,n+1}}{K^{(22)}_{n\,n}}T_{n+1,\,n}^{(2)}
  -\frac{K^{(12)}_{n,\,n}\,K^{(21)}_{n,\,n+1}}
  {K^{(11)}_{n,\,n}\,K^{(22)}_{n,\,n}}T_{n+1,\,n}^{(1)}-\qquad \qquad \qquad \qquad \\[4mm]
  \qquad \qquad \qquad \qquad \qquad
  -\left( \frac{K^{(22)}_{n-1,\,n}}{K^{(22)}_{n-1,\,n-1}}
  -\frac{K^{(12)}_{n,\,n}\,K^{(21)}_{n-1,\,n}}
  {K^{(11)}_{n,\,n}\,K^{(22)}_{n-1,\,n-1}}\right)T_{n,\,n-1}^{(2)},
  \end{array}
\end{equation}
\begin{equation}\label{b2n}
  b^{(2)}_n = \frac{K^{(22)}_{n,\,n}}{K^{(22)}_{n-1,\,n-1}}T_{n,\,n-1}^{(2)},
\end{equation}

\begin{equation}\label{un}
u_n = \frac{K^{(21)}_{n,\,n+1}}{K^{(11)}_{n,\,n}}T_{n+1,\,n}^{(1)}
-\frac{K^{(21)}_{n-1,\,n}\,K^{(22)}_{n,\,n}}
{K^{(11)}_{n,\,n}\,K^{(22)}_{n-1,\,n-1}}T_{n,\,n-1}^{(2)},
\end{equation}

\begin{equation}\label{vn}
v_n = \frac{K^{(12)}_{n,\,n}}{K^{(22)}_{n-1,\,n-1}}T_{n,\,n-1}^{(2)}
-\frac{K^{(12)}_{n-1,\,n-1}\,K^{(11)}_{n,\,n}}
{K^{(22)}_{n-1,\,n-1}\,K^{(11)}_{n-1,\,n-1}}T_{n,\,n-1}^{(1)}.
\end{equation}

\newpage
\begin{table}[ht]
  \centering
  \caption{The eigenvalues and the eigenvector components of the
           Hamiltonian matrix (\ref{Hmat}).}\label{T1}
\begin{ruledtabular}
\begin{tabular}{ccccc}
 $j$ && $\lambda_j$ & $Z_{N,j}$ & $Z_{\mathcal{N},j}$ \\
\hline
 1 & \begin{tabular}{c}
     (a)\\
     (b)\\
    \end{tabular}
& \begin{tabular}{c}
     -0.58970264603\\
     -0.58970263847\\
    \end{tabular}
& \begin{tabular}{c}
     -0.012333185006\\
     -0.012333362733\\
    \end{tabular}
& \begin{tabular}{c}
     0.00018915595025\\
     0.00019663447693\\
    \end{tabular}
    \\
\hline
 2 && 0.40533438179 & 0.18715853083    & 0              \\
 3 && 0.78492505414 & 0.090561490976   & 0              \\
 4 && 1.5244162751 & 0.32069508364    & 0.068124022185 \\
 5 && 1.6473065387 & 0.14429242884    & -0.17954224586   \\
 6 && 2.6728511248 & 0.056915520013 & 0.34337676573    \\
 7 && 3.2807264076 & 0.50263953170    & -0.052996407816\\
 8 && 4.3839439596 & 0.050257472502 & 0.49798274116    \\
\hline
 9 & \begin{tabular}{c}
     (a)\\
     (b)\\
    \end{tabular}
& \begin{tabular}{c}
     6.3135416565\\
     6.3156804888\\
    \end{tabular}
& \begin{tabular}{c}
     0.75676372010\\
     0.75741514694\\
    \end{tabular}
& \begin{tabular}{c}
     -0.062001763734\\
     -0.048160786961\\
    \end{tabular}
    \\
\hline
 10 & \begin{tabular}{c}
     (a)\\
     (b)\\
    \end{tabular}
& \begin{tabular}{c}
     7.5694793828\\
     7.5651078745\\
    \end{tabular}
& \begin{tabular}{c}
     -0.043006216886\\
     -0.029379475185\\
    \end{tabular}
& \begin{tabular}{c}
     -0.76848969549\\
     -0.76948110211\\
    \end{tabular}
    \\
\end{tabular}
\end{ruledtabular}
\end{table}

\newpage
\begin{table}[ht]
  \centering
  \caption{Convergence of $a^{(1)}_{N-1}$,
$a^{(2)}_{N-1}$, $u_{N-1}$ on the left-hand side of (\ref{a12uN}).
  $i$ is the number of iterations.}\label{T2}
\begin{ruledtabular}
\begin{tabular}{cccc}
 $i$ & $a_{N-1}^{(1)}$ & $a_{N-1}^{(2)}$ & $u_{N-1}$\\
\hline
 0 &4.689928491 &5.966326902 &0.0191266184 \\
 1 &4.689911469 &5.965705556 &0.0071475853\\
 2 &4.689912701 &5.965663226 &0.0059643414\\
 3 &4.689912839 &5.965658934 &0.0058474387\\
 4 &4.689912852 &5.965658510 &0.0058359055\\
 5 &4.689912854 &5.965658464 &0.0058347978\\
\end{tabular}
\end{ruledtabular}
\end{table}

\newpage
\begin{table}[ht]
  \centering
  \caption{The elements of the Hamiltonian matrix (\ref{Hmat}).}\label{T3}
\begin{ruledtabular}
\begin{tabular}{ccccccc}
 $n$ & $a_n^{(1)}$ & $b_n^{(1)}$ & $a_n^{(2)}$ & $b_n^{(2)}$ & $u_n$ & $v_n$ \\
\hline
\multicolumn{7}{c}{(a)}\\
\hline
 0 &0.1775666649 & &0.06753394526&  &0.6517927001  & \\
 1 &1.212316541  & -0.3078456533 &2.378081035&-0.01876213797 &0.03459618331 &0.1112824167  \\
 2 &2.233751540  & -0.8637975100 &3.353299231&-0.7967451984  &0.02226447163 &-0.06006980906\\
 3 &3.357567064  & -1.370400686  &4.556450725&-1.319787140   &0.04861624566 &-0.07835689367\\
 4 &4.689928491  & -2.007758312  &5.966326902&-2.032477517   &0.0191266184  &-0.06589047115\\
\hline
\multicolumn{7}{c}{(b)}\\
\hline
 0 &0.2553490884 &             &-0.01023393318&              &0.6404835124 & \\
 1 &1.212396865  &-0.3006058502&2.378394045   &-0.01965488685&0.03937726990& 0.1295217295 \\
 2 &2.233837107  &-0.8638658833&3.353460226   &-0.7967777718 &0.02976290436&-0.06151639336\\
 3 &3.357902488  &-1.370738931 &4.553912267   &-1.319081324  &0.04174355881&-0.07598084688\\
 4 &4.689912854  &-2.008020573 &5.965658464   &-2.031439617  &0.0058347978 &-0.04479790470\\
\end{tabular}
\end{ruledtabular}
\end{table}

\newpage
\begin{figure*}[ht]
\centerline{\psfig{figure=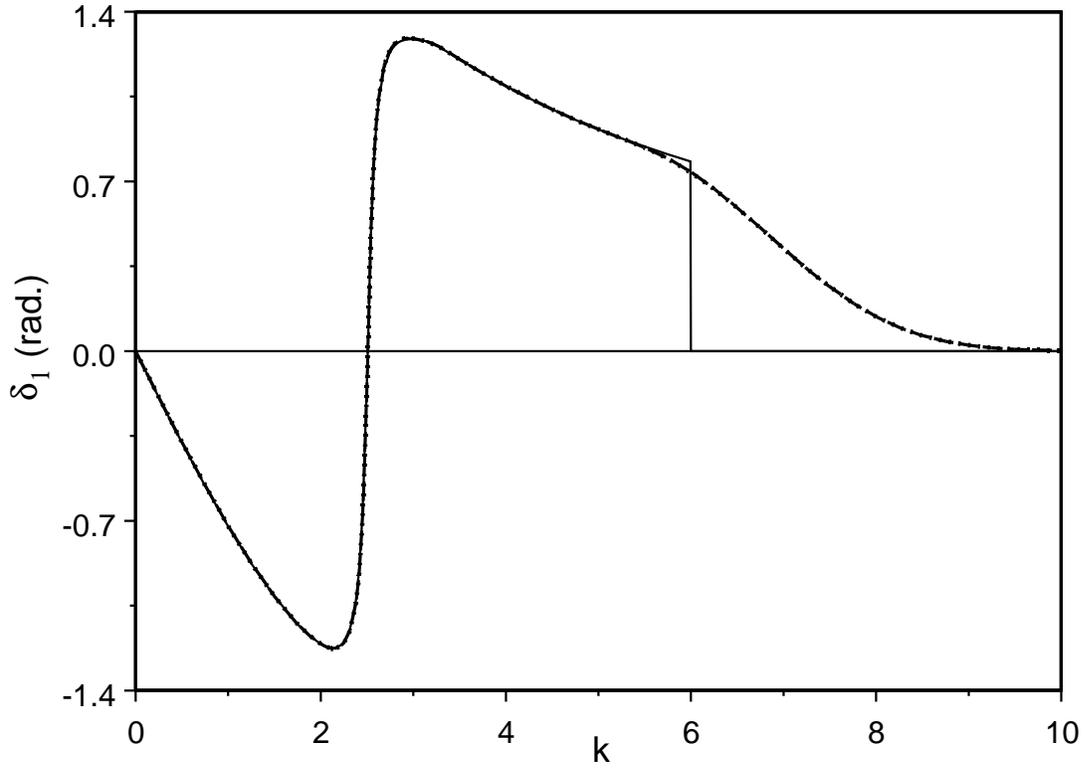,width=1\textwidth}}
\caption{The eigenphase shift $\delta_1$ corresponding to the matrix
${\bf S}(k)$ on the right-hand side of (\ref{Qab}) (thin solid line)
and the resulting matrices: ${\bf S}^{(0)}(k)$ (dashed line) and
${\bf S}^{(5)}(k)$ (dotted line).}
\end{figure*}

\newpage
\begin{figure*}[ht]
\centerline{\psfig{figure=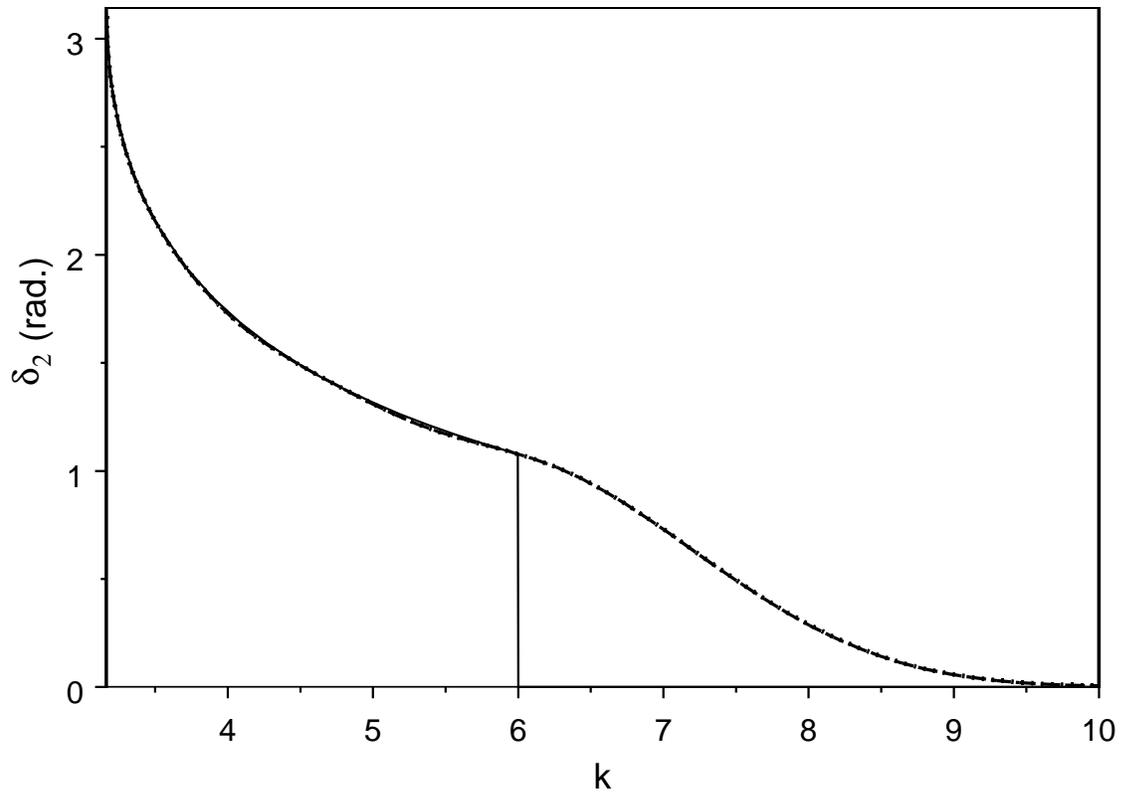,width=1\textwidth}}
\caption{The eigenphase shift $\delta_2$. See Fig.~1 for details.}
\end{figure*}

\newpage
\begin{figure*}[ht]
\centerline{\psfig{figure=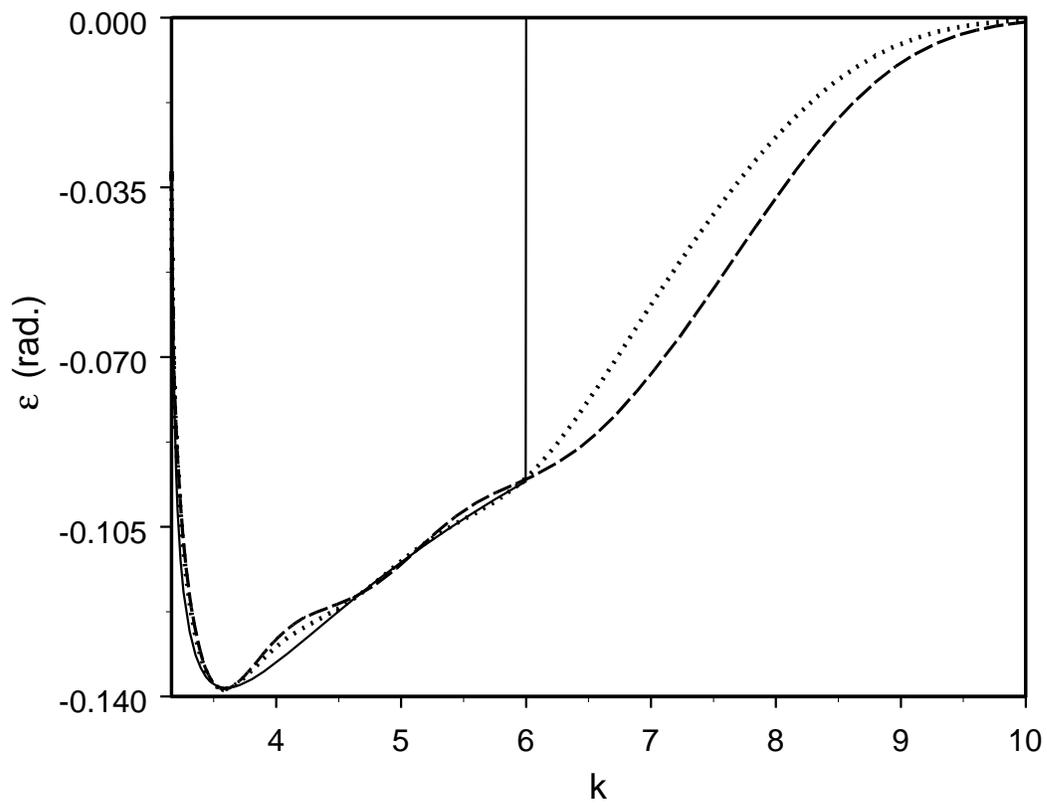,width=1\textwidth}}
\caption{The mixing parameter $\varepsilon$. See Fig.~1 for
details.}
\end{figure*}

\end{document}